\newif\ifpr

\ifpr
\documentclass[amsmath, amssymb, aps, prl, superscriptaddress, twocolumn, floatfix, 10pt]{revtex4-2}
\newcommand{\arxivPR}[2]{#2}
\else
\documentclass[amsmath, amssymb, aps, prl, superscriptaddress, twocolumn, floatfix, 10pt, nofootinbib]{revtex4-1}
\newcommand{\arxivPR}[2]{#1}
\fi
\usepackage[utf8]{inputenc}

\usepackage{gensymb,amssymb,graphicx,color}
\usepackage{dsfont} 
\usepackage{hyperref}
\usepackage[normalem]{ulem}
\usepackage[caption=false]{subfig} 

\makeatletter
\renewcommand\onecolumngrid{
\do@columngrid{one}{\@ne}%
\def\set@footnotewidth{\onecolumngrid}
\def\footnoterule{\kern-6pt\hrule width 1.5in\kern6pt}%
}
\makeatother

\newcommand{\secref}[1]{Sec.\,\ref{#1}}
\newcommand{\refcite}[1]{Ref.\,\cite{#1}}
\newcommand{\refscite}[1]{Refs.\,\cite{#1}}
\newcommand{\eqnref}[1]{Eq.\,\eqref{#1}}
\newcommand{\eqsref}[1]{Eqs.\,\eqref{#1}}
\newcommand{\figref}[1]{Fig.\,\ref{#1}}
\newcommand{\figsref}[1]{Figs.\,\ref{#1}}

\newcommand{\appref}[1]{Appendix\,\ref{#1}}
\newcommand{\appsref}[1]{Appendices\,\ref{#1}}
\newcommand{\bibfoot}[1]{\arxivPR{$^{\text{\cite{#1}}}$}{\cite{#1}}}

\DeclareMathOperator{\tr}{tr}
\DeclareMathOperator*{\Exp}{\text{\scalebox{1.2}{$\mathbb{E}$}}}
\DeclareMathOperator*{\tExp}{\mathbb{E}}

\definecolor{kspink}{RGB}{200,0,200}

\newcommand{\rhoi}{\rho^\text{init}}
\newcommand{\rhof}{\rho^\text{fin}}

\newcommand{\EmQM}{\texorpdfstring{E\lowercase{m}QM}{EmQM} }

\usepackage{mathtools,xparse}

\DeclarePairedDelimiterX{\bra}[1]{\langle}{\rvert}{#1\,}
\DeclarePairedDelimiterX{\ket}[1]{\lvert}{\rangle}{\,#1}
\DeclarePairedDelimiterX{\makebraket}[1]{\langle}{\rangle}{#1}
\NewDocumentCommand{\braket}{som}{%
  \begingroup\activatebraketbar
  \IfBooleanTF{#1}
    {\makebraket*{#3}}
    {\IfNoValueTF{#2}{\makebraket{#3}}{\makebraket[#2]{#3}}}%
  \endgroup
}
\makeatletter
\newcommand{\braketbar}{%
  \,\delimsize\vert\@ifnextchar|{\!}{\,}%
}
\makeatother
\newcommand{\activatebraketbar}{%
  \begingroup\lccode`~=`|\lowercase{\endgroup\let~}\braketbar
  \mathcode`|="8000
}

\hypersetup{
  colorlinks = true,
  urlcolor = blue,
  pdfauthor = {Kevin Slagle},
  pdftitle = {Slagle - 2022 - Testing Quantum Mechanics using Noisy Quantum Computers}
}

\begin{document}

\title{Testing Quantum Mechanics using Noisy Quantum Computers}

\author{Kevin Slagle}
\affiliation{Department of Electrical and Computer Engineering, Rice University, Houston, Texas 77005 USA}
\affiliation{Department of Physics and Institute for Quantum Information and Matter, California Institute of Technology, Pasadena, California 91125, USA}
\affiliation{Walter Burke Institute for Theoretical Physics, California Institute of Technology, Pasadena, California 91125, USA}


\begin{abstract}
We outline a proposal to test quantum mechanics in the high-complexity regime using
  noisy intermediate-scale quantum (NISQ) devices.
The procedure involves simulating a non-Clifford random circuit, followed by its inverse,
  and then checking that the resulting state is the same as the initial state.
We are motivated by the hypothesis that quantum mechanics is not fundamental,
  but instead emerges from a theory with less computational power,
  such as classical mechanics.
This emergent quantum mechanics (EmQM) hypothesis makes the prediction that quantum computers will not be capable of sufficiently complex quantum computations.
We show that quantum mechanics predicts that the fidelity of our procedure decays exponentially with circuit depth
  (due to noise in NISQ devices),
  while EmQM predicts that the fidelity will decay significantly more rapidly for sufficiently deep circuits,
  which is the experimental signature that we propose to search for.
We estimate rough bounds for when possible signals of EmQM should be expected.
Furthermore, we find that highly informative experiments should require only thousands qubits.
\end{abstract}

\maketitle

Quantum mechanics has enjoyed roughly a century of predictive success.
Although there have been many high-precision tests of quantum mechanics \cite{experimentRev1992,testQM2000,photonTests,SticklerTesting},
  it is possible that a new experimental regime could discover violations of quantum mechanics.
That is, quantum theory might merely be an approximation to a more precise underlying theory.
In order to be consistent with past experiments,
  perhaps deviations from quantum mechanics are only noticeable in the high-complexity regime,
  i.e. the regime of highly entangled states that require deep quantum circuits to prepare.
The complexity of a wavefunction can be defined as the number of gates (\emph{e.g.} 2-qubit unitaries)
  required to make the wavefunction from a product state
  \cite{BrownSusskindComplexity,SusskindComplexity,BrandaoComplexity,NicoleComplexity}.
The high-complexity regime has remained relatively untested since decoherence and noise
  make it tremendously difficult to probe this regime with high fidelity.
The advent of quantum computers introduces an exciting opportunity to test quantum mechanics \cite{SadanaTestQuantum} in the high-complexity regime for the first time.
We would like to obtain evidence for whether quantum mechanics makes correct predictions at large complexity scales
  using a protocol that requires as few quantum computing resources as possible.
In this work, we outline an efficient protocol that is appropriate for noisy intermediate-scale quantum (NISQ) \cite{NISQ} devices
  or quantum computers with only limited error correction.

Leading tests of quantum mechanics in the high-complexity regime
  include the recent quantum advantage experiments \cite{GoogleSupremacy,ZuchongzhiAdvantage,ZuchongzhiAdvantage2,PhotonAdvantage}.
The superconducting qubit experiments \cite{GoogleSupremacy,ZuchongzhiAdvantage,ZuchongzhiAdvantage2} performed up to 24 layers of random gates on up to 60 qubits,
  followed by measurements in the computational basis.
The infidelity of their circuits can be predicted from, and was verified to be consistent with,
  the infidelity of the individual gates in the low-complexity regime.
This helps verify quantum mechanics up to the complexity regime of 60 qubits entangled by depth 24 circuits.
However, since the ideal measurement outcomes of their random circuits can not be efficiently calculated on classical computers,
  it is impractical to extend this methodology to deeper circuits with more qubits.
To overcome this, we propose to apply the inverse of the quantum circuit before taking measurements so that the ideal measurement outcome is trivial to calculate (similar to simple circuit mirroring \cite{Proctor2008}).

Our primary motivation is to obtain evidence for or against the hypothesis that
  quantum mechanics emerges from a local classical model.
We define a local classical model to be any lattice model that is classical (i.e. not quantum) with local dynamics.
That is, a local classical model consists of a lattice, where the state at each lattice site is defined by a finite list of numbers (which does not grow with the system size), and the time evolution of each lattice site only depends on the state of nearby sites.
We allow the time evolution to be stochastic and either discrete or continuous.
Cellular automata \cite{WolframNKS,Hooft,HooftOntology} and local classical lattice Hamiltonians (and Lagrangians) are examples of local classical models.
Although there have been many preliminary attempts to construct a model of emergent quantum mechanics (EmQM)
  \cite{EmQM,SlaglePreskill,Hooft,HooftBook,Adler,VanchurinEntropic,VanchurinWorld,VanchurinNeural,Nelson2012},
  it is still not clear to what extent an EmQM model could be consistent with previous experiments.
We are also motivated by recent progress in tensor network \cite{OrusTN,IsometricTN,2dDMRG,canonicalPEPS,tensorMC}
  and neural network \cite{MendlNeural,SchmittNeural2D,PollmannNeural,JonssonNN} algorithms
  and the possibility that future algorithms might yield a model of EmQM.
Note that many local classical models are not local hidden variable theories,
  which implies that they are not ruled out by Bell experiments
  (in addition to the possibility that hidden information could travel significantly faster than light \cite{SlaglePreskill});
  see \appref{app:Bell} for further discussion.

In order to motivate a strategy for detecting deviations from quantum mechanics,
  we recall that local classical models can be efficiently simulated on classical computers
  while classical computers (very likely) can not efficiently simulate quantum mechanics \cite{PostIQP,AaronsonSupremacy,CharacterizingSupremacy}.
Therefore, models of EmQM must show discrepancies from quantum theory
  at sufficiently high complexity \cite{HooftComputers,HooftBook}.
However, if quantum mechanics emerges at \emph{e.g.} the Planck scale,
  then an EmQM model could effectively have at its disposal the computational power of
  one CPU core per Planck volume operating at a Planck frequency.
That is an incredible amount of computational power.
As such, it is conceivable that a local classical model
  could exhibit EmQM well enough to be in agreement with previous experiments,
  but not sufficiently complex quantum computations,
  such as the experiment we propose.

\section{Proposal}
\begin{figure}
  \centering
  \includegraphics[width=\columnwidth]{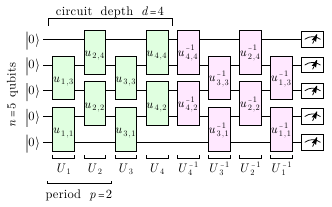} \\
  \caption{%
    Our protocol: repeatedly apply a random circuit of a random depth $d$ to $n$ qubits, followed by its inverse,
      and then measure in the computational basis.
    For each random sample, the 2-qubit gates $u_{t,i}$ are chosen randomly.
    The depth-dependent fidelity $F_d$ is the fraction of depth-$d$ samples with final state $\ket{0}$.
    Although we depict a circuit with one-dimensional connectivity for simplicity,
      higher-dimensional connectivity is preferable.
  }\label{fig:circuit}
\end{figure}

The protocol, which we summarize in \figref{fig:circuit}, is to repeat the following:
\begin{enumerate}
  \item initialize the zero state $\ket{0}$ on $n$ qubits
  \item apply $d$ layers of random gates $U_1 \cdots U_d$,
        with depth $d$ randomly chosen between 0 and $d_\text{max}$
  \item apply the inverse circuit $U_d^\dagger \cdots U_1^\dagger$
  \item measure in the computational basis and check if the final state is the zero state $\ket{0}$
\end{enumerate}
The max depth $d_\text{max}$ should be significantly larger than the depth required
  to strongly entangle all pairs of qubits.
That is, $d_\text{max} \gg n^{1/D}$ for a circuit with $D$-dimensional connectivity of the 2-qubit gates
  (\emph{e.g.} $d_\text{max} \approx 10 \, n^{1/D}$),
  or $d_\text{max} \gg \log_2 n$ for all-to-all connectivity.
Each layer of gates $U_t$ is a product of 2-qubit gates,
  which are independently and randomly chosen for each sample and layer.
The random 2-qubit gates can be chosen from any distribution
  (of at least two gates\footnote{%
    Using random instead of fixed gates reduces fluctuations in the exponential fidelity decay
      that result from coherent errors \cite{randomizedCompiling}.}),
  as long as the entire forward circuit $U_1 \cdots U_d$ is expected to be difficult to simulate classically.
We require that the random distribution for each layer of gates repeats with a small period $p$;
  \emph{i.e.} $U_t$ and $U_{t+p}$ are independently chosen from the same probability distribution.
The depth-dependent fidelity $F_d$ is the fraction of depth-$d$ samples for which the measured final state $\ket{0}$ is measured.
The protocol is complete once the fidelities $F_d$ have been measured to the desired statistical precision.

To make the best use of each sample,
  each depth $d$ can be sampled with probability $\sim F_d^{-2}$.
The circuit depth $d$ is chosen randomly for each sample in order to mitigate
  the effect of slow drifts in the gate fidelity between samples.
(See \appref{app:drift} for more on drift.)

\begin{figure}
  \centering
  \includegraphics[width=.7\columnwidth]{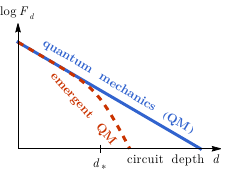} \\
  \caption{%
    Exponential fidelity decay $F_d \sim e^{-\lambda d}$ with circuit depth $d$ predicted by quantum mechanics (QM) (blue)
      vs a schematic sketch of the emergent quantum mechanics (EmQM) expectation (red).
    EmQM agrees with QM up to a critical depth $d_*$,
      after which the EmQM fidelity must start decaying significantly faster
      (assuming sufficiently large $n$ and small $\lambda$).
  }\label{fig:decay}
\end{figure}

We show that quantum mechanics predicts that the fidelity should decay exponentially $F_d \sim e^{-\lambda d}$
  with the circuit depth $d$ (until saturation near $F_\infty \sim 2^{-n}$)
  for Markovian noise models.
In contrast, if quantum mechanics emerges from a theory with significantly less computational power,
  such as classical mechanics,
  and if $\lambda$ is sufficiently small
  to be in the quantum advantage regime \cite{StoudenmireLimits},
  then we argue that the fidelity should start to decay significantly faster after a sufficiently large depth $d_*$,
  as depicted in \figref{fig:decay}.

Our protocol is similar to randomized benchmarking \cite{noiseEstimation,tdesignRB,twirlRB,twirlRBlong,GateRB,KnillRB,RBGeneral,randomCircuitSampling}
  and uses a similar mathematical foundation.
One difference is that we allow arbitrary non-Clifford gates so that the quantum advantage regime can be reached,
  while randomized benchmarking typically focuses on Clifford gates so that the inverse unitary can be a low-depth circuit.
The low-depth inverse circuit also differs from our proposal since we instead apply a long sequence of inverted unitaries in the reverse order
  (as in a Loschmidt echo \cite{LoschmidtEcho}).
The long sequence of inverted unitaries is typically avoided in randomized benchmarking since it can hide errors
  that are important to quantify for benchmarking quantum computers \cite{practicalVerification,Proctor2008}.
However, this is not an issue for testing quantum mechanics since we are not interested in detecting such subtle errors;
  we are instead interested in more catastrophic errors resulting from deviations from quantum mechanics.
Another difference is that randomized benchmarking typically studies a small number of qubits,
  while we focus on many $n \gg 1$ qubits.
Conveniently, we find that a larger number of qubits helps make the quantum mechanics prediction for the fidelity decay be much closer to an exponential decay.

\section{\EmQM Expectation}
\label{sec:EmQM}
We expect that an EmQM model would not have enough computational power
  to accurately perform the forward circuit
  (green in \figref{fig:circuit}) of our protocol
  for sufficiently many qubits $n$ and large depth $d$.
In particular, we assume that if the zero state $\ket{0}$ is measured at the end of the protocol,
  then an EmQM would have had to simulate the circuit well enough such that the EmQM could also
  randomly sample bit strings from the forward circuit
  (\emph{i.e.} bit strings $b$ from the probability distribution $P(b) = |\braket{b|U_d \cdots U_1|0}|^2$).
This is the same sampling task considered in recent quantum advantage experiments \cite{GoogleSupremacy,ZuchongzhiAdvantage,ZuchongzhiAdvantage2}
  and is expected to require an exponential amount of CPU time for classical algorithms \cite{circuitSupremacy,MovassaghSupremacy,BoulandSupremacy,AaronsonXEB,optimalSA}.
Building on these assumptions, we can roughly predict when our protocol might see signals of EmQM.

Consider the brute-force Schr\"{o}dinger algorithm (SA) \cite{SchrodingerStyle},
  which involves simply applying unitary matrices to a $2^n$-component wavefunction.
This algorithm can simulate a depth $d$ circuit of $n$ qubits
  in $N_\text{CPU}^\text{SA} \approx n d \, 2^n \sim 2^n$ CPU cycles.
We are not aware of any faster algorithms that could sample bit strings from random non-Clifford circuits
  with large depth $d \gg n^{1/D} \gg 1$ for $D$-dimensional circuit connectivity (or $d \gg \log n \gg 1$ for all-to-all connectivity) and
  large gate fidelity\footnote{%
    Gate fidelity above the error-correction threshold $f \approx 0.99$ \cite{RaussendorfThreshold,RaussendorfThresholdMCQB}
      is presumably sufficient.}
  (even if only a small circuit fidelity $F \sim 10^{-3}$ is required)
  \cite{circuitSupremacy,ZuchongzhiAdvantage,AaronsonSupremacy,optimalSA}.

Consider a (perhaps hypothetical) local classical EmQM model where quantum mechanics emerges at length and time scales $L_\text{EmQM}$ and $T_\text{EmQM}$.
Now consider the computational cost required to classically simulate the EmQM model,
  which is describing the EmQM of a quantum computer performing the aforementioned circuit.
Suppose that a quantum computer runs the circuit within length and times scales
  $L_\text{QPU}$ and $T_\text{QPU}$.
The computational cost should then be roughly 
  $N_\text{CPU}^\text{EmQM} \sim \frac{T_\text{QPU}}{T_\text{EmQM}} \left(\frac{L_\text{QPU}}{L_\text{EmQM}}\right)^{D_\text{EmQM}}$.
$D_\text{EmQM}$ is the number of spatial dimensions of the EmQM model,
  which could be greater than three if the EmQM model hosts extra spatial dimensions \cite{SlaglePreskill}.
(More cautiously, we could take $L_\text{QPU}$ be the spatial length required for the EmQM model to capture the quantum computer's operation of the circuit.)

Assume that the EmQM model does not yield a faster algorithm to classically simulate deep random quantum circuits than the Schr\"{o}dinger algorithm.
With this assumption,
  we can bound $N_\text{CPU}^\text{SA} \lesssim N_\text{CPU}^\text{EmQM}$.
Next we solve for $n$ in $2^n \sim N_\text{CPU}^\text{SA} \lesssim N_\text{CPU}^\text{EmQM}$
  to roughly upper bound the largest number of qubits $n_*$ that an EmQM model could accurately handle at large depth:
\begin{align}
  n_* &\lesssim \log_2 \frac{T_\text{QPU}}{T_\text{EmQM}} + D_\text{EmQM} \log_2 \frac{L_\text{QPU}}{L_\text{EmQM}} \label{eq:n crit}\\
      &\stackrel{\text{e.g.}}{\sim} 130 + \log_2 \frac{T_\text{P}}{T_\text{EmQM}} + D_\text{EmQM} \left(110 + \log_2 \frac{L_\text{P}}{L_\text{EmQM}} \right) \nonumber
\end{align}
Therefore when $n \geq n_*$, there should be a critical random circuit depth $d_*$
  beyond which the EmQM model should deviate from quantum mechanics predictions (as in \figref{fig:decay}).
If this were not the case, then the EmQM model would yield an algorithm more efficient than the Schr\"{o}dinger algorithm for random circuit simulation,
  which we assumed to not be the case.
As an example, we express the second line relative to Planck length $L_\text{P}$ and time $T_\text{P}$ scales
  with $L_\text{QPU} \sim 1$\,centimeter and $T_\text{QPU} \sim 10^{-3}$\,seconds.

\eqnref{eq:n crit} is remarkably predictive because huge changes in any of the length or time scales
  only modestly affect our bound on $n_*$ due to the logarithmic dependence.
In particular,
  only thousands of qubits should be needed to significantly rule out the EmQM hypothesis.
For example, increasing the QPU scales to $T_\text{QPU} = 100$\,years and $L_\text{QPU} = 100$\,light-years in the second line only
  increases our estimated bound on $n_*$ by $42 + 66 D_\text{EmQM}$.
Or if we assume $D_\text{EmQM} \leq 6$, $T_\text{EmQM} \geq \varepsilon\, T_\text{P}$, and $L_\text{EmQM} \geq \varepsilon\, L_\text{P}$,
  then we obtain the bounds $n_* \lesssim 800$, $n_* \lesssim 2000$, or $n_* \lesssim 5400$
  if we assume that $\varepsilon = 1$, $\varepsilon = 10^{-50}$, or $\varepsilon = 10^{-200}$, respectively.
It is also incredible that linearly increasing the qubit count $n$ facilitates the ability to test the EmQM hypothesis down to exponentially small length and time scale!

In order to bound $n_*$ using the Schr\"{o}dinger algorithm argument,
  we had to assume $d \gg n^{1/D} \gg 1$ (or $d \gg \log n \gg 1$ for all-to-all connectivity).
Therefore, we can roughly bound the critical depth $d_* \lesssim n_*^{1/D}$ (or $d_* \lesssim \log n$ for all-to-all).
In an attempt to safely exceed this bound,
  below we consider max circuit depths
\begin{equation}
  d_\text{max} \approx \begin{cases} 10\, n^{1/D}  & \text{for $D$-dimensional circuits} \\
                                     10\, \log_2 n & \text{for all-to-all circuits} \end{cases} \label{eq:d}
\end{equation}
  for which it seems likely that $d_\text{max} > d_*$.
However, it might be useful to also probe larger depths.

With $N_\text{s} = 10^8$ samples,
  the fidelity can be measured down to
  $F_\text{min} \approx N_\text{s}^{-1/2} \epsilon_\text{r}^{-1} \approx 10^{-3}$
  with a relative error of $\epsilon_\text{r} \approx 10\%$.
We will show that if each 2-qubit gate has an entanglement fidelity\bibfoot{Ffoot:gateFidelity} $f$,
  then quantum mechanics predicts that our protocol has fidelity $F_d \sim f^{n d}$,
  where $n d$ is the approximate number of 2-qubit gates.
Therefore, statistically significant data can be obtained only for sufficiently small quantum volumes:
\begin{equation}
  n d_\text{max} \lesssim \frac{\ln F_\text{min}}{\ln f} \approx \frac{\ln F^{-1}_\text{min}}{1-f} \label{eq:volume}
\end{equation}

Recall that the larger $n$ is, the better we can test the EmQM hypothesis (which requires $n \gtrsim n_*$),
  as long as we can measure the fidelity of sufficiently deep circuits with reasonably small relative statistical error.
Thus, for a given $F_\text{min}$ (determined by the number of measurement samples)
  gate fidelity $f$,
  we can solve \eqsref{eq:volume} and \eqref{eq:d}
  for the max number of qubits $n$ and max depth $d_\text{max}$.

For example, for circuits with gate infidelity $1-f = 10^{-3}$ and $F_\text{min} = 10^{-3}$,
  statistical significance requires $n d_\text{max} \lesssim 7000$ [\eqnref{eq:volume}].
For a circuit with $D=2$ dimensional connectivity,
  \eqnref{eq:d} then implies that we should use circuits with
  roughly $n \approx 80$ qubits and max depth $d_\text{max} \approx 90$
  (which is significantly larger than the $n\approx60$ and $d\approx24$ used in \refscite{GoogleSupremacy,ZuchongzhiAdvantage,ZuchongzhiAdvantage2}).
If we instead consider random circuits of 2-qubit gates with all-to-all coupling (rather than nearest neighbor),
  then we use the second case of \eqnref{eq:d},
  which allows the max depth $d_\text{max} \approx 70$ to be slightly smaller
  and the qubit count $n \approx 100$ to be slightly larger.

With an improved gate infidelity $1-f = 0.8 \times 10^{-5}$ (which may require error correction techniques \cite{GottesmanErrorCorrection})
  and $F_\text{min} = 10^{-3}$,
  statistical significance allows much larger quantum volumes $n d \lesssim 8.6 \times 10^5$.
For $D=2$ dimensional connectivity,
  this allows $n \approx 2000$ qubits with max depth $d_\text{max} \approx 440$,
  which could rule out the EmQM hypothesis with higher confidence.
If we have access to all-to-all connected circuits and want to be able to run the protocol with $n \approx 2000$ qubits (and $F_\text{min} = 10^{-3}$),
  then the gate infidelity could be about four times larger than the $D=2$ case:
  $1-f \lesssim 3 \times 10^{-5}$.

\section{Quantum Mechanics Prediction}
\label{sec:decay}
We now argue that quantum mechanics predicts that the fidelity $F_d$ decays exponentially with circuit depth $d$
  under physically reasonable noise models.
Intuitively, each gate $u$ has a probability $p_u$ of making an error,
  resulting in a gate entanglement fidelity $f_u \approx 1-p_u$.
The circuit fidelity $F_d$ is then roughly the probability $f^{n d}$ that none of the $\approx n d$ two-qubit gates made an error:
\begin{equation}
  F_d \approx \widetilde{F}_0 \, e^{-\lambda d} \label{eq:F}
\end{equation}
  where $\lambda \approx - n \ln f$ and $f$ is an average gate entanglement fidelity
  (until the fidelity eventually saturates near $F_\infty \sim 2^{-n}$).
In the experiment, $\widetilde{F}_0$ and $\lambda$ are fitting parameters.
We substantiate this intuition below.
In \appref{app:twirl}, we provide additional evidence
  that the fidelity should decay exponentially (according to quantum mechanics)
  using a 2-design twirling argument;
  see also \cite{mirrorTheory,Dalzell2021}.

As a starting approximation, we model the fidelity using gate-dependent but time-independent noise:
\begin{equation}
  F_d = \Exp_{U_1, \ldots, U_d}
    \tr\!\big[\rhof \, (\Phi_{U_1^\dagger} \!\circ\cdots\circ \Phi_{U_d^\dagger} \!\circ \Phi_{U_d} \!\circ\cdots\circ \Phi_{U_1}) (\rhoi)\big] \label{eq:F model}
\end{equation}
The initial state is modeled by a density matrix $\rhoi$.
Measuring if the final state is the zero state
  is modeled by the quantum channel $\rho \to \tr(\rhof \, \rho)$.
$\rhoi$ and $\rhof$ should be close to the $n$-qubit zero state $\ket{0} \bra{0}$,
  up to some infidelity resulting from state preparation and measurement (SPAM) errors.
$\tExp_{U_1, \ldots, U_d}$ averages over the allowed layers of random gates.
The quantum channel $\Phi_{U_t}$ models the noisy quantum computer implementation
  of a particular layer of random gates $U_t$.

As depicted in \figref{fig:circuit},
  each layer of gates $U_t$ is a product of 2-qubit gates $U_t = \prod_i u_{t,i}$,
  where $i$ labels a pair of qubits.
We similarly model the quantum channel $\Phi_{U_t}$ as
  a composition of noisy quantum channels:
  $\Phi_{U_t} = \circ_i \Phi_{u_{t,i}}$.
Each 2-qubit channel $\Phi_u$ can be decomposed as $\Phi_u(\rho) = \mathcal{N}_u(u \rho u^\dagger)$,
  where the noise channel $\mathcal{N}_{u}$ implicitly depends on the unitary $u$ and the two qubits that it acts on.

We expand the noise channel for each gate using a Pauli decomposition\footnote{%
    $n$-qubit noise channels could be used instead to model cross-talk \cite{crosstalk} or cosmic ray \cite{GoogleCosmicRay} errors
      with minimal changes to our arguments.}:
\begin{equation}
  \mathcal{N}_u(\rho) = \sum_{\alpha,\beta \in \{0,1,2,3\}^2} \chi^{(u)}_{\alpha\beta} \, \sigma^\alpha \, \rho \, \sigma^\beta
\end{equation}
  where $\chi^{(u)}$ is a positive semi-definite matrix and
  $\sigma^\alpha \in \{I, X, Y, Z\}^{\otimes 2}$ is a $2$-qubit Pauli operator.
The gate entanglement fidelity\bibfoot{Ffoot:gateFidelity}
  $f_u = \chi^{(u)}_{\mathbf{00}}$ plays an important role
  since all other terms in the expansion result in an error and very small fidelity.
The terms with $\alpha\neq{\mathbf 0}$ or $\beta\neq{\mathbf 0}$ can be thought of as resulting from discrete errors.
For large gate fidelities ($f_u \approx 1$),
  errors are suppressed by small
  $|\chi^{(u)}_{\alpha\beta \neq \mathbf{00}}| \leq \sqrt{1-f_u}$.\bibfoot{Cfoot:chiBound}
Therefore multiple errors will typically be separated by many gates,
  which will spread out \cite{NahumSpreading,KeyserlingkSpreading,SusskindSpreading,approximateTDesign,Dalzell2021}
  any non-identity $\sigma^\alpha$ or $\sigma^\beta$ factor into a highly non-local operator.
After spreading out, the contribution to the fidelity in \eqnref{eq:F model}
  resulting from one or more errors is very small $O(2^{-n})$.\bibfoot{Dfoot:errorFidelity}
If there are many $n \gg 1$ qubits,
  any error will thus typically result in a negligible contribution to the fidelity.

Therefore, the fidelity $F_d$ in \eqnref{eq:F model} is roughly the fraction of times that no error occurs.
$F_0 = \tr \rhof \rhoi$ is the probability that a SPAM error occurred.
The probability that no error occurs during the forward and backward
  2-qubit gates $u_{t,i}$ and $u_{t,i}^\dagger$ is
\begin{equation}
  f_i^2 = \Exp_{u_{t,i}} f_{u_{t,i}} f_{u_{t,i}^\dagger} \label{eq:fi}
\end{equation}
  where $\Exp_{u_{t,i}}$ averages over the allowed 2-qubit gates.
Therefore, the probability that no errors occur over the entire circuit of $2n_\text{g} d/p$ gates is roughly
\begin{align}
  F_d &\approx F_0 f^{2 n_\text{g} d/p} &\text{with}&&
  f^{2n_\text{g}} &= \prod_{i=1}^{n_\text{g}} f_i^2 \label{eq:f}
\end{align}
  where $f$ is an average gate entanglement fidelity
  and $n_\text{g}$ is the number of gates in a period (\emph{e.g.} $n_\text{g}=4$ in \figref{fig:circuit}).
We thus obtain \eqnref{eq:F} with
\begin{align}
  \widetilde{F}_0 &\approx F_0 = \tr \rhof \rhoi &
  \lambda &\approx - \frac{2n_\text{g}}{p} \ln f \label{eq:lambda}
\end{align}
  where $\lambda$ reduces to $\lambda \approx - n \ln f$ when $n_\text{g} \approx p n/2$, as is typically the case.

\begin{figure}
  \centering
  \includegraphics[width=.85\columnwidth]{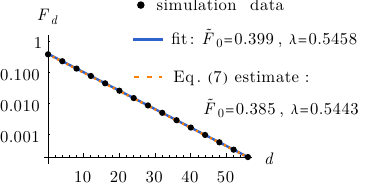} \\
  \caption{%
    Simulation data (black dots) showing the quantum mechanics prediction of
      the exponential fidelity $F_d$ decay
      for a $5\times5$ grid of $n=25$ noisy qubits and Sycamore/Zuchongzhi-like \cite{GoogleSupremacy,ZuchongzhiAdvantage} gates.
    Also shown is a fit (blue line) to an exponential decay [\eqnref{eq:F}]
      and the rough predictions (dotted orange line) from \eqnref{eq:lambda}.
  }\label{fig:simulation}
\end{figure}

Although \eqnref{eq:lambda} is only an approximate expression for $\widetilde{F}_0$ and $\lambda$ since multiple errors can sometimes cancel out
  (especially if they are spatially close or at mirrored positions in the circuit),
  we numerically find that \eqnref{eq:lambda} is a very good approximation for 
  many qubits $n \gg 1$ and large gate fidelity $f \approx 1$.
(See \appref{app:twirl} for a more accurate expression.)
As an example, in \figref{fig:simulation} we show a simulation for a $5\times5$ grid of $n=25$ qubits
  using roughly Sycamore/Zuchongzhi-like \cite{GoogleSupremacy,ZuchongzhiAdvantage} gates with gate-dependent errors
  and average gate entanglement fidelity $f \approx 0.993$ using a Markovian noise model \eqnref{eq:F model}.
(See \appref{app:numerics} for details.)

\section{Example: \EmQM from MPS}
\label{sec:MPS}
It is illuminating to consider the time evolution of matrix product states (MPS) \cite{MPS} with fixed bond dimension
  as an (incomplete) toy model of emergent quantum mechanics (EmQM) in one spatial dimension.
We do not consider MPS algorithms to be a legitimate model of EmQM because
  it is currently not known how to generalize MPS algorithms to a three-dimensional local classical model
  that could be compatible with previous experiments (see \appref{app:MPS});
  although there has been remarkable related algorithmic progress \cite{IsometricTN,2dDMRG,canonicalPEPS,ParallelDMRG}.

\begin{figure}
  \centering
  \includegraphics[width=.7\columnwidth]{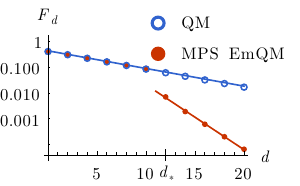} \\
  \caption{%
    Simulation data (red dots) showing the circuit fidelity $F_d$ decay of a matrix product state (MPS) toy model of EmQM
      with bond dimension $\chi = 2^{10}$ for a chain of 24 noisy qubits.
    Similar to \figref{fig:decay},
      the toy EmQM model agrees with the quantum mechanics (QM) prediction (blue circles)
      until a critical depth $d_* = 12$,
      after-which the fidelity decays at a significantly faster exponential rate.
    Lines are fits to these two regimes.
  }\label{fig:MPS}
\end{figure}

In \figref{fig:MPS}, we show a MPS simulation
  of our protocol with noisy gates (similar to \figref{fig:simulation})
  for a chain of 24 qubits.
(See \appref{app:MPS simulation} for details.)
During the MPS simulation, bond dimensions are truncated to $\chi \leq 2^{10}$,
  which limits the amount of R\'enyi-0 entanglement entropy \cite{RenyiMPS} across any cut of the chain.
The maximum entanglement is reached at the critical depth $d_*$,
  after which the fidelity decay is dominated by the truncation error \cite{StoudenmireLimits} of the MPS algorithm (instead of the gate infidelity)
  and the fidelity $F_d$ begins to decay at a faster exponential rate.
This is precisely the EmQM signature depicted in \figref{fig:decay}.

\section{Discussion}

Our protocol for testing quantum mechanics has several unique and essential properties:
(1) It is remarkably simple and appropriate for NISQ devices or devices with minimal error correction.
(2) It can distinguish between violations of quantum mechanics and noise.
(3) It makes efficient use of quantum volume in pursuit of searching for deviations from quantum mechanics in the high complexity regime.
In particular, the first half of gates are used to generate a high-complexity state as quickly as possible.
In this sense, only the second half of the circuit is overhead necessary for checking the result.

A disadvantage of our protocol is that we had to assume that the hypothetical EmQM theory would deviate significantly from quantum mechanics during the first half of the circuit,
  even though the full circuit is easy to compute.
In this sense, our protocol can not rule out all hypothetically possible EmQM models.
To overcome this limitation, it could be useful to investigate how cryptographic techniques used for quantum certification and verification \cite{FalsifiableQM,interactiveProofs,verificationOverview,verification,MahadevVerification} could be applied without adding significant overhead to the quantum volume.
Although we expect that most natural (i.e. not overly contrived) EmQM models will not be able to take advantage of this loophole,
  more research on EmQM is needed to understand this possibility better.

If an exponential fidelity decay (blue in \figref{fig:decay})
  is observed for high-depth circuits on many qubits,
  then our arguments in \secref{sec:EmQM} will significantly rule out possible theories of emergent quantum mechanics (EmQM),
  including \refscite{Hooft,HooftBook,Adler,VanchurinEntropic,VanchurinWorld,VanchurinNeural,Nelson2012,SlaglePreskill}.
It may also be possible to rule out other modifications \cite{Wolfram2020,InteractingWorlds,discreteProbabilityZee,discreteHilbertZee,PalmerDiscretised,Torrome} to quantum mechanics.
Quantum certification and verification techniques \cite{interactiveProofs,verificationOverview,verification,FalsifiableQM,MahadevVerification}
  or Shor's algorithm \cite{Shor,Shor2048} could rule out deviations from quantum mechanics with even higher confidence,
  although this would require less noisy quantum computers with more qubits due to additional overhead.

If a signal suggestive of EmQM (red in \figref{fig:decay}) is experimentally observed,
  then future work will be required
  to either determine what the true underlying theory is
  or rule out possible explanations consistent with quantum mechanics
  (\emph{e.g.} non-Markovian noise or state leakage \cite{limitsRB,correlationsRB}).

If quantum mechanics does emerge from a local classical model,
  then the computational power of quantum computers could be severely limited \cite{HooftComputers,HooftBook}.
For example, BQP-hard problems would only be tractable for limited problem sizes.
But quantum computers would still be incredibly useful technology,
  particularly as a device to probe new fundamental physics.

However, it is possible that violations of quantum mechanics could make quantum computers more powerful for certain tasks.
For example, if it is possible to violate the probabilistic Born's rule such that postselection could be performed,
  then quantum computers may be able to efficiently solve problems in the PostBQP = PP \cite{PostBQP,PostIQP} complexity class up to a bounded problem size.
This could be possible if the EmQM prefers to drop branches of the wavefunction with a complexity that is too high\footnote{%
    To exploit this, consider a quantum computer in a superposition of orthogonal states $\ket{\psi_1} + \ket{\psi_2}$
      consisting of many qubits ($n > n_*$) but complexity below the critical complexity.
    Then maybe one could postselect the $\ket{\psi_1}$ state by applying a controlled high-depth random unitary $U$
      to only the $\ket{\psi_2}$ component:
      \unexpanded{$\ket{\psi_1} + \ket{\psi_2} \underset{\text{controlled-} U}{\overset{\text{apply}}{\to}}
                   \ket{\psi_1} + U \ket{\psi_2} \underset{\text{drops } U \ket{\psi_2}}{\overset{\text{EmQM}}{\to}} \ket{\psi_1}$.}}
  (which would be a possible mechanism behind the EmQM signature of rapid fidelity decay).

\begin{acknowledgments}
We thank Alex Dalzell, Andru Gheorghiu, Ulysse Chabaud, Sean Carroll, Spiros Michalakis, Vitaly Vanchurin, Leonid Pryadko, Xie Chen, Natalie Klco, Frank Pollmann, Shivaji Sondhi, Curt von Keyserlingk, Itai Arad, and Tsz Chun Wu
  for helpful conversations.
We also thank John Preskill for detailed comments.
K.S. is supported by the Walter Burke Institute for Theoretical Physics at Caltech; and
  the U.S. Department of Energy, Office of Science, National Quantum Information Science Research Centers, Quantum Science Center.
\end{acknowledgments}

\bibliography{testingQuantum}

\begin{thebibliography}{107}
\bibitem [{Ffo()}]{Ffoot:gateFidelity}%
  \BibitemOpen
  \href@noop {} {}\bibinfo {note} {The average fidelity of a $m$-qubit noise
  channel $\mathcal{N}$ acting on a normalized random $n$-qubit state is \\
  $\displaystyle f^{(n)} = \Exp_{\psi \in \mathbb{C}^{2^n}} \braket[\big]{\psi|
  \mathcal{N}\big(\ket{\psi}\bra{\psi}\big) |\psi} = \frac{f + 2^{-n}}{1 +
  2^{-n}}$ \\ $f$ is the average entanglement fidelity
  \cite{NielsenEntanglementFidelity,SchumacherEntanglementFidelity} and the $n
  \to \infty$ limit of $f^{(n)}$: \\ $\displaystyle f = \lim_{n \to \infty}
  f^{(n)} = 2^{-2m} \sum_k \tr(E_k) \tr(E_k^\dagger)$ \\ where
  $\mathcal{N}(\rho) = \sum_k E_k \rho E_k^\dagger$ is the Kraus decomposition
  of $\mathcal{N}$. The average gate fidelity is often defined as
  $f^{(m)}$.}\BibitemShut {Stop}%
\bibitem [{Cfo()}]{Cfoot:chiBound}%
  \BibitemOpen
  \href@noop {} {}\bibinfo {note} {The diagonal terms are bounded by
  $\big|\chi^{(u)}_{\alpha\alpha}\big| \leq 1-f_u$ for $\alpha \neq {\mathbf
  0}$ since $\mathcal{N}_u$ is trace preserving: $\tr \mathcal{N}_u(\rho) = \tr
  \rho$. The absolute value of the off-diagonal $\alpha \neq \beta$ terms are
  then upper bounded by $\big|\chi^{(u)}_{\alpha\beta}\big| \leq
  \sqrt{\big|\chi^{(u)}_{\alpha\alpha}\big| \big|\chi^{(u)}_{\beta\beta}\big|}
  \leq \sqrt{f_u \, (1-f_u)}$ since $\chi^{(u)}$ is a positive semi-definite
  matrix.}\BibitemShut {Stop}%
\bibitem [{Dfo()}]{Dfoot:errorFidelity}%
  \BibitemOpen
  \href@noop {} {}\bibinfo {note} {After spreading out, a single Pauli error
  results in a fidelity $\sim \Exp_{\psi} |\braket{\psi | \sigma | \psi}|^2 =
  (2^n + 1)^{-1}$, where $\psi$ and $U$ are random $n$-qubit states and
  unitaries and $\sigma$ is a Pauli operator.}\BibitemShut {Stop}%
\end{thebibliography}

\onecolumngrid
\newpage
\appendix






\section{Exponential Fidelity Decay via 2-design Twirling}
\label{app:twirl}

Given a gate independent noise assumption and 2-design\footnote{%
  We thank Alex Dalzell for pointing out that a 2-design approximation is sufficient, rather than a 4-design.}
  approximation,
  we can generalize the randomized benchmarking 2-design twirling argument \cite{tdesignRB} to our protocol.
Similar to the main texts, we find that quantum mechanics predicts an exponential fidelity decay [\eqnref{eq:f}].

We assume that the noise is gate and time independent,
  but it may depend on the position of the 2-qubit gate.
Therefore, in the notation of \secref{sec:decay},
  the noise channels are assumed to be the same for different 2-qubit gates $u_{t,i}$ and $u_{t',i}$
  acting on the same pair of qubits labeled by $i$:
  $\mathcal{N}_{u_{t,i}} = \mathcal{N}_{u_{t',i}}$.
However, we can allow the noise for the forward and backward circuits to be different.
As is also the case in the main text,
  it is not essential to assume that the noise only comes from 2-qubit error channels.

We also approximate each layer of 2-qubit gates $U_t$ as a 2-design.
Formally, this approximation is not justified
  since an approximate 2-design requires a $O(n^{1/D})$-depth $D$-dimensional circuit \cite{approximateTDesign}.
However, this approximation might be better than expected if we can instead think of it as
  only approximating the composition of $O(n^{1/D})$ layers of gates as a 2-design.
We do not consider this possibility here
  since it would complicate the definition of the noise channel $\mathcal{N}$ below.

To simplify the notation, we rewrite the Markovian fidelity \eqnref{eq:F model} from the main text as
\begin{equation}
  F_d = \Exp_{U_1, \ldots, U_d} \tr( \rho'_d \, \rho_d )
\end{equation}
  where
\begin{gather}
\begin{aligned}
  \rho'_d &\equiv (\Phi_{U_d^\dagger}^\dagger \!\circ\cdots\circ \Phi_{U_1^\dagger}^\dagger) (\rhof) \\
  \rho_d  &\equiv (\Phi_{U_d} \!\circ\cdots\circ \Phi_{U_1}) (\rhoi)
\end{aligned}
\end{gather}
  implicitly depend on $U_1, \ldots, U_d$, and
  $\Phi^\dagger$ is defined such that $\tr[\Phi^\dagger(\rho') \, \rho] = \tr[\rho' \, \Phi(\rho)]$.
Below we recursively integrate out one $U_t$ at a time by
  approximating each $U_t$ as an arbitrary Haar random unitary (or a unitary 2-design) on $n$ qubits:
\begin{align}
  F_d &= \Exp_{U_1, \ldots, U_d} \tr\!\left[ \mathcal{N}_d'^{\dagger}(\rho'_{d-1}) \, U_d^\dagger \mathcal{N}_d(U_d \rho_{d-1} U_d^\dagger) U_d \right] \label{eq:F1}\\
      &\approx \frac{\hat{F}_d-2^{-2n}}{1-2^{-2n}} \Exp_{U_1, \ldots, U_{d-1}} \tr\!\left[ \mathcal{N}_d'^\dagger(\rho'_{d-1}) \, \rho_{d-1} \right] + (1-\hat{F}_d) \frac{2^{-n}}{1-2^{-2n}} \tr \mathcal{N}_d'^\dagger(\rho'_{d-1}) \label{eq:F2}\\
      &\approx \left( \prod_{t=d'+1}^d \frac{\hat{F}_t-2^{-2n}}{1-2^{-2n}} \right) \Exp_{U_1, \ldots, U_{d'}}
         \tr\!\left[ \mathcal{N}_{d'}'^\dagger(\rho'_{d'-1}) \, U_{d'}^\dagger (\mathcal{N}'_{d'+1} \circ \mathcal{N}_{d'})(U_{d'} \rho_{d'-1} U_{d'}^\dagger) U_{d'} \right] \\
      &\quad\quad + \sum_{t=d'+1}^d (1-\hat{F}_t) \frac{2^{-n}}{1-2^{-2n}} \tr \mathcal{N}_t'^\dagger(\rho'_{t-1}) \nonumber\\
      &\approx \left( \prod_{t=1}^d \frac{\hat{F}_t-2^{-2n}}{1-2^{-2n}} \right) 
         \tr\!\left[ \mathcal{N}_1'^\dagger(\rhof) \, \rhoi \right] + \sum_{t=1}^d (1-\hat{F}_t) \frac{2^{-n}}{1-2^{-2n}} \tr \mathcal{N}_t'^\dagger(\rho'_{t-1}) \\
      &= \tr\!\left[ \mathcal{N}_1'^\dagger(\rhof) \, \rhoi \right] \, \prod_{t=1}^d \hat{F}_t + O(2^{-n}) \label{eq:F5}
\end{align}
$\mathcal{N}_t$ and $\mathcal{N}'_t$ are gate-independent noise channels of the forward and backward circuits:
  $\Phi_{U_t}(\rho) = \mathcal{N}_t(U_t \rho U_t^\dagger)$ and
  $\Phi_{U_t^\dagger}(\rho) = \mathcal{N}'_d(U_t^\dagger \rho U_t)$.
In the last line, we drop very small $O(2^{-n})$ terms.
$\hat{F}_t$ is the entanglement fidelity of $\mathcal{N}'_{t+1} \circ \mathcal{N}_t$ for $t=1,\ldots,d-1$,
  while $\hat{F}_d$ is just the entanglement fidelity of $\mathcal{N}_d$:
\begin{align}
  \hat{F}_t &= 2^{-2n} \tr\!\left(\mathcal{N}'_{t+1} \circ \mathcal{N}_t\right) &
  \hat{F}_d &= 2^{-2n} \tr \mathcal{N}_t \label{eq:hatF}\\
      &= 2^{-2n} \sum_{k',k} \left|\tr\!\left(E_{k'}'^{(t+1)} E_k^{(t)}\right)\right|^2 &
      &= 2^{-2n} \sum_{k} \left|\tr E_k^{(t)}\right|^2 \nonumber
\end{align}
$E_k^{(t)}$ and $E_{k'}'^{(t)}$ are Kraus operators of $\mathcal{N}_t$ and $\mathcal{N}'_t$; \emph{i.e.}
  $\mathcal{N}_t(\rho)  = \sum_k    E^{ (t)}_k    \rho\, E^{(t)\dagger}_k$ and
  $\mathcal{N}'_t(\rho) = \sum_{k'} E'^{(t)}_{k'} \rho\, E'^{(t)\dagger}_{k'}$.
Note that $\hat{F}_t = \hat{F}_{t+p}$ has periodicity $p$ since the circuit geometry is assumed to have periodicity $p$.

The entanglement fidelity of the composition $\mathcal{N}'_{t+1} \circ \mathcal{N}_t$ appears in \eqnref{eq:hatF}
  because coherent errors can cancel out.
For example, if $\mathcal{N}_t(\rho) = V \rho V^\dagger$ and $\mathcal{N}'_{t+1}(\rho) = V^\dagger \rho V$ for some unitary $V$,
   then the fidelity $F_d$ would not depend on the coherent error $V$.
This subtlety does not occur in randomized benchmarking \cite{noiseEstimation,tdesignRB,twirlRB}
  because randomized benchmarking uses a compactified inverse circuit.

\eqnref{eq:F5} is an exponential fidelity decay
  $F_d \approx \widetilde{F}_0 e^{-\lambda d}$
  with parameters
\begin{gather}
\begin{aligned}
  \widetilde{F}_0 &\approx \tr\!\left[ \mathcal{N}_1'^\dagger(\rhof) \, \rhoi \right] \frac{\tr \mathcal{N}_p}{\tr\!\left(\mathcal{N}'_{p+1} \circ \mathcal{N}_p\right)} \\
  \lambda &\approx -\frac{1}{p} \log \prod_{t=1}^p \hat{F}_t
\end{aligned} \label{eq:lambdaApp}
\end{gather}
  when $d$ is an integer multiple of the period $p$.

\eqnref{eq:lambdaApp} differs from \eqnref{eq:lambda} in the main text for two reasons:
  \eqnref{eq:lambda} neglected the fact that coherent errors in $\mathcal{N}'_{t+1} \circ \mathcal{N}_t$ can cancel out,
  while \eqnref{eq:lambdaApp} assumes that errors are gate-independent.
Unfortunately, the decay rate $\lambda$ in \eqnref{eq:lambdaApp} can not be written in terms of a product of 2-qubit gate fidelities
  [as in \eqnref{eq:lambda}]
  because $\mathcal{N}'_{t+1} \circ \mathcal{N}_t$ in the definition of $\hat{F}_t$
  involves two circuit layers with different positions of the 2-qubit gates.
If we were to make the rough approximation
  $2^{-2n} \tr\!\left(\mathcal{N}'_{t+1} \circ \mathcal{N}_t\right) \to
    2^{-4n} \tr\mathcal{N}'_{t+1} \tr \mathcal{N}_t$,
  then $\lambda$ in \eqnref{eq:lambdaApp} would simplify to \eqnref{eq:lambda}.

\section{Numerical Simulations}
\label{app:numerics}

In \figref{fig:simulation} we showed a simulation for a $5\times5$ grid of $n=25$ qubits
  using roughly Sycamore/Zuchongzhi-like \cite{GoogleSupremacy,ZuchongzhiAdvantage,ZuchongzhiAdvantage2} gates with gate-dependent errors.
\appsref{app:gates} and \ref{app:noise} specify the gate and noise models that we used.
\appref{app:Kraus} defines an optimized basis of Kraus operators that we used to minimize the amount of statistical sampling error.
All plots in the main text have statistical errors that are too small to be seen.
\appref{app:MPS simulation} specifies details used to produce the MPS data in \figref{fig:MPS}.

\begin{figure}
  \centering
  \subfloat[]{\includegraphics[width=.48\columnwidth]{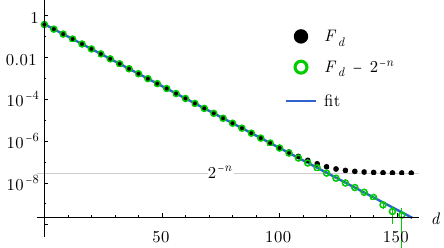}} $\quad$
  \subfloat[\label{fig:simulationAppErr}]{\includegraphics[width=.48\columnwidth]{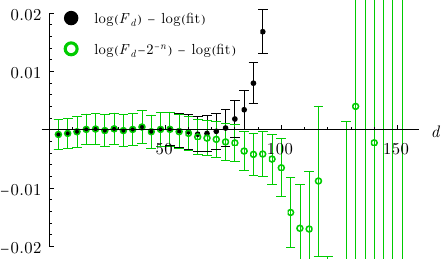}} \\
  \subfloat[\label{fig:circuit2d}]{\includegraphics[width=.6\columnwidth]{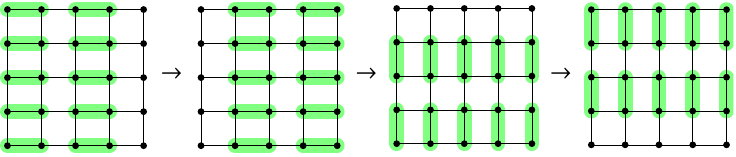}}
  \caption{%
    {\bf (a)} The same simulation data (black dots) as \figref{fig:simulation} (but for larger max depth $d_\text{max}$)
      showing the quantum mechanics prediction of
      the exponential fidelity $F_d$ decay
      for a $5\times5$ grid of $n=25$ noisy qubits.
    The fidelity $F_d$ saturates to $2^{-n}$ at large circuit depth $d$.
    Subtracting $2^{-n}$ from $F_d$ (green circles) leads to a continued exponential decay
      (at least until our simulation data becomes dominated by statistical errors).
    Also shown is a fit (blue line) to an exponential decay.
    {\bf (b)} With $2^{-n}$ subtracted, the fit residuals are dominated by statistical errors for our simulation of 10,000 Monte Carlo samples.
    Error bars denote one standard deviation (\emph{i.e.} 68\% confidence intervals).
    {\bf (c)} The period $p=4$ ordering of 2-qubit gates (green) applied to the $5\times5$ array of qubits.
  }\label{fig:simulationApp}
\end{figure}

In \figref{fig:simulationApp}, we show a more detailed version of \figref{fig:simulation}
  showing a larger max circuit depth $d_\text{max}$.
At large depth, the fidelity $F_d$ saturates to $2^{-n}$,
  which will be too small to be experimentally observable for large $n \gtrsim 50$.
\figref{fig:simulationAppErr} shows the residual errors to the exponential decay fit.
The $\log(F_d) - \text{fit}$ residuals are dominated by the $2^{-n}$ saturation at large depth.
After subtracting off $2^{-n}$, we see that $F_d - 2^{-n}$ agrees with an exponential decay up to statistical errors.
The statistical errors of $F_d - 2^{-n}$ become very large once the fidelity decays to $2^{-n}$
  due to the $2^{-n}$ subtraction.

\begin{figure}
  \centering
  \subfloat[]{\includegraphics[width=.48\columnwidth]{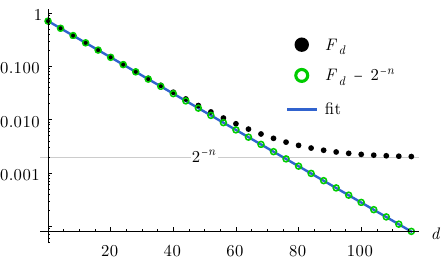}} $\quad$
  \subfloat[\label{fig:simulationAppErr3x3}]{\includegraphics[width=.48\columnwidth]{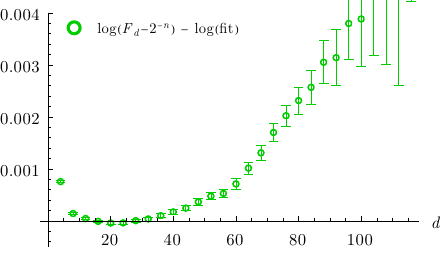}}
  \caption{%
    Same as \figref{fig:simulationApp}, but for a $3\times3$ grid of $n=9$ qubits,
      for which we obtained $8\times10^7$ Monte Carlo samples.
    Panel (b) shows that the residual errors of $F_d - 2^{-n}$ to an exponential fit are very small.
  }\label{fig:simulationApp3x3}
\end{figure}

To investigate the residuals with less statistical error,
  we also ran a simulation with less qubits
  (for which we can run significantly more Monte Carlo samples).
\figref{fig:simulationApp3x3} is similar to \figref{fig:simulationApp},
  but for a simulation of a $3\times3$ grid of $n=9$ qubits
  and a smaller average gate entanglement fidelity $f \approx 0.987$ (instead of $f \approx 0.993$).
In \figref{fig:simulationAppErr3x3},
  we see that residual errors to an exponential fit are very small when $F_d \gg 2^{-n}$.

\subsection{Gate Model}
\label{app:gates}

In our numerical simulations,
  we use gates that are similar to the Sycamore/Zuchongzhi gates \cite{GoogleSupremacy,ZuchongzhiAdvantage,ZuchongzhiAdvantage2}.
The gates are applied in the order shown in \figref{fig:circuit2d}.

Each 2-qubit gate $u_{t,i}$ is an iSWAP gate
  (rather than Sycamore/Zuchongzhi iSWAP-like gates \cite{GoogleSupremacy,ZuchongzhiAdvantage,ZuchongzhiAdvantage2})
\begin{equation}
  U_\text{iSWAP} = \begin{pmatrix}
                     1 &  0 &  0 & 0 \\
                     0 &  0 & -i & 0 \\
                     0 & -i &  0 & 0 \\
                     0 &  0 &  0 & 1
                   \end{pmatrix}
\end{equation}
  that is preceded by a pair of randomly chosen 1-qubit gates ($\sqrt{X}$, $\sqrt{Y}$, or $\sqrt{W}$) on two qubits,
  where $W = \frac{1}{\sqrt{2}} (X+Y)$.
Using $U_\text{iSWAP}^\dagger = U_\text{iSWAP} \, (Z \otimes Z)$,
  we decompose the inverse gate $u_{t,i}^\dagger$ as an iSWAP gate followed by a
  $\sqrt{X}^\dagger Z$, $\sqrt{Y}^\dagger Z$, or $\sqrt{W}^\dagger Z$ gate on each of the two qubits.

\subsection{Noise Model}
\label{app:noise}

In our numerical simulations,
  we use a generic error model with gate and position dependent errors.
Each iSWAP gate for each pair of qubits (labeled by $i$) has a unique noise channel.
And for each of the $n$ qubits, we apply a unique noise channel for each of six the 1-qubit gates
  ($\sqrt{X}$, $\sqrt{Y}$, $\sqrt{W}$, $\sqrt{X}^\dagger Z$, $\sqrt{Y}^\dagger Z$, and $\sqrt{W}^\dagger Z$).
Each noise channel is generated randomly.
None of our plots appear to differ qualitatively for different random instances of gate error channels.

One way to generate a random noise channel on a $d$-dimensional Hilbert space is to
  choose a set of $m$ Kraus operators as a Haar random $md \times d$ semi-unitary matrix $E_{ka,b}$
  (\emph{i.e.} $\sum_{ka} E_{ka,b'}^* E_{ka,b} = \delta_{b',b}$).
Then the noise channel is
\begin{equation}
  \mathcal{N}_m(\rho) \overset{\text{rand}}{=} \sum_{k=1}^m E_k \rho E_k^\dagger
\end{equation}
In order to cover the set of all noise channels, one must choose $m=d^2$.
However, for $d=4$ or larger, this leads to a distribution of noise channels that is
  strongly concentrated near the depolarizing channel.
In order to obtain a broader but still generic distribution of noise channels,
  we generate each random noise channel as an average of the above random noise channel for $m=1,2,\ldots,d^2$:
\begin{equation}
  \mathcal{N}_\text{rand}(\rho) \overset{\text{rand}}{=} d^{-2} \sum_{m=1}^{d^2} \mathcal{N}_m(\rho)
\end{equation}

We also apply a randomly-chosen coherent error for each gate.
For a $d$-dimensional Hilbert space, we generate coherent error channels using a random unitary that is generated by
\begin{equation}
  U_p \overset{\text{rand}}{=} e^{i \sqrt{\frac{p}{d-1}} H^\text{Gauss}_\text{rand}}
\end{equation}
  where $H^\text{Gauss}_\text{rand}$ is a complex standard Gaussian random matrix
  and $p$ tunes the infidelity.

Each noise channel is then generated via
\begin{equation}
  \mathcal{N}_p(\rho) \overset{\text{rand}}{=} U_{p''/2} \left[ \left(1-\tfrac{p'}{2}\right) \rho + \tfrac{p'}{2} \, \mathcal{N}_\text{rand}(\rho) \right] U_{p''/2}^\dagger \label{eq:noise}
\end{equation}
  where $p'$ and $p''$ are randomly chosen between $p/\sqrt{2}$ and $\sqrt{2} \,p$.
The entanglement fidelity of $\mathcal{N}_p$ is roughly $1-p$.

We randomly generated the 2-qubit iSWAP noise channel using $\mathcal{N}_{p_2}$,
  while the 1-qubit noise channels were randomly generated using $\mathcal{N}_{p_1}$,
  with $p_1 = 0.1 p_2$.
For \figsref{fig:simulation}, \ref{fig:MPS}, and \ref{fig:simulationApp}, we used $p_2 = 0.005$.
For \figref{fig:simulationApp3x3}, we used $p_2 = 0.01$.

\subsection{Simulation Algorithm Details}
\label{app:Kraus}

We simulate the fidelity in \eqnref{eq:F model} by Monte Carlo (MC) sampling the density matrices and quantum channels
  such that only two wavefunctions are stored in memory: one from $\rhoi$ and the other from $\rhof$.
To reduce the amount of statistical error,
  we sample the density matrices and quantum channels such that each MC sample has the same average entanglement fidelity.

That is, $\rhoi$ and $\rhof$ are uniformly sampled from unnormalized wavefunctions
  $\ket{\psi^\text{init}_k}$ and $\ket{\psi^\text{fin}_k}$:
\begin{equation}
\begin{aligned}
  \rhoi &= \Exp_k \, \ket{\psi^\text{init}_k} \bra{\psi^\text{init}_k} \\
  \rhof &= \Exp_k \, \ket{\psi^\text{fin}_k} \bra{\psi^\text{fin}_k}
\end{aligned} \label{eq:wavefunctions}
\end{equation}
  which are chosen to have the same fidelity:
\begin{equation}
\begin{aligned}
  \braket{0|\psi^\text{init}_k} &= \braket{0|\psi^\text{init}_{k'}} \\
  \braket{0|\psi^\text{fin}_k} &= \braket{0|\psi^\text{fin}_{k'}}
\end{aligned} \label{eq:wavefunctions constraint}
\end{equation}
In \eqnref{eq:wavefunctions}, $\tExp_k \equiv n_k^{-1} \sum_k$
  denotes an expectation value over $n_k$ uniformly random $k$.
Similarly, the quantum channels $\Phi_u$ are rewritten using a Kraus decomposition
\begin{equation}
  \Phi_u(\rho) = \Exp_k E^{(u)}_k \rho \, E^{(u)\dagger}_k
\end{equation}
  (with nonstandard normalization due to the $\Exp_k$)
  such that each Kraus operator has the same trace:
\begin{equation}
  \tr E^{(u)}_k = \tr E^{(u)}_{k'} \label{eq:Kraus constraint}
\end{equation}
This implies that each Kraus operator has the same entanglement fidelity as $\Phi_u$:
  $f_u = 2^{-2m} \tr E^{(u)}_k \tr E^{(u)\dagger}_k$
  where $E^{(u)}_k$ is a $2^m \times 2^m$ matrix.
With this optimization,
  for a given sequence of unitaries $U_1,\ldots,U_d$,
  each MC sample has the approximately the same fidelity,
  which leads to a significant reduction in MC statistical error.

We only require that these special wavefunctions and Kraus operators obey linear constraints
  [\eqnref{eq:wavefunctions constraint} and \eqref{eq:Kraus constraint}].
They can therefore be obtained from another decomposition (\emph{e.g.} an eigendecomposition),
  $\rho = \tExp_{\tilde{k}} \ket{\widetilde{\psi}_{\tilde{k}}} \bra{\widetilde{\psi}_{\tilde{k}}}$ or
  $\Phi(\rho) = \tExp_{\tilde{k}} \widetilde{E}_{\tilde{k}} \rho \widetilde{E}^\dagger_{\tilde{k}}$,
  via a unitary transformation $V_{k,\tilde{k}}$ that rotates the vector
  $v_{\tilde{k}} = \braket{0|\widetilde{\psi}_{\tilde{k}}}$ (or $v_{\tilde{k}} = \tr \widetilde{E}_{\tilde{k}}$)
  to $\sum_{\tilde{k}} V_{k,\tilde{k}} v_{\tilde{k}} = \sqrt{f}$
  for some constant fidelity $f>0$.
That is, $\ket{\psi_k} = V_{k,\tilde{k}} \ket{\widetilde{\psi}_{\tilde{k}}}$
  and $E_k = V_{k,\tilde{k}} \widetilde{E}_{\tilde{k}}$.

Each MC sample for a sequence of fidelities $F_d$ with $d=0,\ldots,d_\text{max}$ is then obtained as follows:
\begin{enumerate}
  \item sample initial forward circuit wavefunction: $\ket{\psi}  \leftarrow \ket{\psi^\text{init}_k}$ for uniformly random $k$
  \item sample final   backward circuit wavefunction: $\ket{\psi'} \leftarrow \ket{\psi^\text{fin}_k}$ for uniformly random $k$
  \item foreach circuit depth $d=1,\ldots,d_\text{max}$:
  \begin{enumerate}
    \item foreach 2-qubit gate position $i$:
    \begin{enumerate}
      \item randomly sample an allowed gate: $u_{d,i}$
      \item apply random Kraus operator for forward gate: $\ket{\psi}  \leftarrow E^{(u_{d,i})}_k \ket{\psi}$
            for uniformly random $k$
      \item apply random Kraus operator for backward gate: $\ket{\psi'} \leftarrow E^{(u_{d,i}^\dagger)\dagger}_k \ket{\psi'}$
            for uniformly random $k$
    \end{enumerate}
    \item obtain fidelity sample: $F_d \leftarrow |\braket{\psi' | \psi}|^2$
  \end{enumerate}
\end{enumerate}

\subsection{MPS Simulation}
\label{app:MPS simulation}

In \figref{fig:MPS}, we showed a MPS simulation of our protocol with noisy gates for a chain of 24 qubits.
We used the same noise and gate models and sampling algorithm as in \figref{fig:simulation},
  \emph{i.e.} the models in \appsref{app:gates} and \ref{app:noise} and algorithm in \appref{app:Kraus}.

Random gates, SPAM errors, and random Kraus operators are randomly sampled
  as described in \appref{app:Kraus}.
However, a subtlety results from the fact that the MPS simulation introduces non-linear truncation errors.
When there is no truncation error, as in the quantum mechanics prediction (blue in \figref{fig:MPS}),
  then changing the Kraus operator basis $E_k \to V_{k,k'} E_{k'}$
  does not affect the fidelity.
But when there is truncation error, the choice of Kraus operator basis will affect the simulation result.
Typically, Kraus operators that are more like measurement operators will reduce the amount of entanglement,
  which decreases the amount of truncation error in later truncation steps.
Alternatively, Kraus operators that are closer to unitary operators
  typically result in a larger truncation error.
We use the Kraus basis defined in \appref{app:Kraus} for the MPS simulation,
  which is a nice choice because all of the Kraus operators behave similarly with similar fidelity.

\section{Drift}
\label{app:drift}

While iterating our protocol many times (\emph{e.g.} over the course of hours)
  to obtain a large number of samples,
  slow temperature fluctuations (and other causes) can result in miscalibrations (and other complications)
  that slowly drift \cite{drift} over time.
As a result, the effective error channel of each gate can change over time,
  resulting in slowly-varying gate fidelities \cite{cycleBenchmarking} for different samples.
To mitigate issues resulting from drift,
  our protocol samples a random circuit depth $d$ for each sample.
If circuit depth was instead sampled in increasing order,
  then drift could potentially cause the gate fidelity to decrease over time,
  which could then lead to a fidelity $F_d$ decay that could be mistaken for the EmQM signature.
By randomly sampling circuit depth,
  we can show that the effect of slow temporal drift
  can only cause the exponential fidelity $F_d$ decay rate to slightly decrease with circuit depth $d$,
  rather than increase as for EmQM.
This occurs because if the drift occurs much slower than the time scale to sample all circuit depths,
  then for larger depth, the fidelity $F_d$ is dominated by rare samples with rare but higher gate fidelities (due to rare drifts),
  while at smaller depth the fidelity is dominated by typical gate fidelities.
As such, the effects of drift are unlikely to be confused with the EmQM signature.

To see this more explicitly, we can consider the following toy model.
Assume that quantum mechanics is correct and
  suppose that temporal drifts occur over a time scale that is much longer than the time it takes to run a single sample in our protocol.
Then suppose that due to drift, the probability that a sample measures a zero state is the following time-dependent fidelity:
\begin{equation}
  F_d(t) = e^{-\lambda(t) d}
\end{equation}
Similar to \eqnref{eq:F},
  the fidelity decays exponentially with circuit depth $d$
  (although we neglect SPAM errors, \emph{i.e.} the $\widetilde{F}_0$ factor, for simplicity).
However, the fidelity also depends on time
  since the exponential decay rate $\lambda(t)$ is time-dependent due to slow time-dependent drifts.

\begin{figure}
  \centering
  \includegraphics[width=.35\columnwidth]{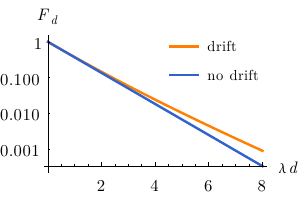} \\
  \caption{%
    An exponential fidelity decay $F_d = e^{-\lambda d}$ without drift (blue)
      vs a fidelity decay with drift (orange) [\eqnref{eq:drift12} with $\lambda_1 = \frac{2}{3}\lambda$ and $\lambda_2 = \frac{4}{3}\lambda$].
    Drift only causes the fidelity to decay slightly slower at larger depth,
      which can not be mistaken for the EmQM signature [shown in \figref{fig:decay}].
  }\label{fig:drift}
\end{figure}

To gain intuition, suppose that the fidelity decay rate $\lambda(t)$ drifts between $\lambda_1$ and $\lambda_2$ with a uniform probability distribution in-between.
Then the average probability that a sample measures a zero state is the average fidelity:
\begin{equation}
\begin{aligned}
  F_d &= \frac{1}{\lambda_2 - \lambda_1} \int_{\lambda_1}^{\lambda_2} e^{-\lambda d} \, \mathrm{d}\lambda \\
    &= \frac{1}{d} \frac{e^{-\lambda_1 d} - e^{-\lambda_2 d}}{\lambda_2 - \lambda_1} 
\end{aligned} \label{eq:drift12}
\end{equation}
See \figref{fig:drift} for an example plot.
At small depth, $F_d \approx e^{-\lambda d}$ where $\lambda = \frac{1}{2} (\lambda_1 + \lambda_2)$,
  while at large depth $F_d \approx (\lambda_2 - \lambda_1)^{-1} \frac{1}{d} \, e^{-\lambda_1 d}$.
Thus the fidelity decays slower at larger depth (in contrast to the EmQM signature)
  since $\lambda_1 < \lambda$.

Now consider a general probability distribution $P(\lambda)$ of fidelity decay rates $\lambda(t)$ after time averaging.
Then the fidelity is $F_d = \int_\lambda e^{-\lambda d}$
  where $\int_\lambda f(\lambda) \equiv \int_0^\infty f(\lambda) P(\lambda) \,\mathrm{d}\lambda$
  averages over the different decay rates.
In this general setting, we can also show that the fidelity decays slower at larger depth.
The exponential decay rate near a given depth is
  the average of $\lambda(t)$:
\begin{equation}
  - \partial_d \log F_d
    = \braket{\lambda}_d
        \quad\text{where}\quad \braket{f(\lambda)}_d \equiv \frac{\int_\lambda f(\lambda) \, e^{-\lambda d}}{\int_\lambda e^{-\lambda d}}
\end{equation}
The slope of the curve $\log F_d$ is thus $-\braket{\lambda}_d$.
The derivative of this slope is then the variance of $\lambda(t)$:
\begin{equation}
  -\partial_d \braket{\lambda}_d
    = \braket[\big]{(\lambda - \braket{\lambda}_d)^2}_d
\end{equation}
Since the above variance is always positive, the derivative of the slope of $\log F_d$ vs $d$ is always positive.
This implies that the fidelity decays slower at larger depth than slower depth,
  which is opposite of the EmQM signature.

\newpage
\section{MPS}
\label{app:MPS}

In \secref{sec:MPS}, we considered the time evolution of matrix product states (MPS) \cite{MPS} with fixed bond dimension
  as an (incomplete) toy model of emergent quantum mechanics (EmQM) in one spatial dimension.
And in \appref{app:realism},
  we explain how and why tensor networks are not ruled out by Bell experiments.
However, we do not consider MPS to be a legitimate model of EmQM because
  it is currently not known how (or if it is possible) to generalize MPS algorithms to a three-dimensional local classical model
  that could be compatible with previous experiments.
We elaborate on this point below.

One of the hallmarks of MPS is the canonical form,
  which allows one to calculate the expectation value of a local operator near some position $x$
  in terms of only the tensors near $x$.
When generalizing MPS to higher dimensions,
  this sense of canonical form can either be retained,
  as is done for isometric tensor networks \cite{IsometricTN,2dDMRG,canonicalPEPS};
  or the canonical form can be dropped,
  as is done for projected entangled pair states (PEPS) \cite{OrusTN};
  or the canonical form can be replaced with something new (but less constraining),
  such as a weighted trace gauge (WTG) \cite{EvenblyCanonical}.

When the canonical form is retained \cite{IsometricTN,2dDMRG,canonicalPEPS},
  a local update algorithm to time evolve the tensor network is known \cite{IsometricTN}.
However, the algorithm does not quite yet yield a local classical model since the update algorithm has to iteratively sweep across the tensor network.
This is similar to many MPS algorithms, such as TDVP \cite{TDVP},
  which time evolve a length-$L$ chain by performing a forward sweep to update the tensors at
  sites $i$ and $i+1$ for $i=1,2,\ldots,L-1$,
  followed by a backward sweep with $i=L-1,L-2,\ldots,1$.
It would be incredibly unnatural if the time evolution of the universe sweeps across space in this serial (rather than parallel) manner.
Therefore, we expect an EmQM model to have the form of a local classical model:
  for which the local updates are performed in parallel for all sites at once.
Nevertheless, we expect that it is possible to parallelize the time-evolving isometric tensor network algorithm \cite{IsometricTN}
  into a continuous local classical model,
  \emph{e.g.} by using ideas from \refcite{ParallelDMRG}.

However, there is a more severe issue that is also shared with MPS:
  between certain pairs of well-separated spatial disks of radius $\ell$,
  isometric tensor networks \cite{IsometricTN} and canonical PEPS \cite{canonicalPEPS,2dDMRG}
  can only represent wavefunctions that have a bounded amount of entanglement between the two regions.
In particular, the mutual information $I$ between certain pairs of regions is bounded by the log of the bond dimension $\chi$:
  $I < \log \chi$.
Even a large bond dimension $\chi = 1024$ is only large enough to encode 10 bell pairs between the two regions.
Unless $\chi$ is extraordinarily large,
  this is incompatible with many experiments, including the recent quantum advantage experiments
  \cite{GoogleSupremacy,ZuchongzhiAdvantage,ZuchongzhiAdvantage2,PhotonAdvantage}.

An EmQM model should allow the mutual information to diverge with the length scale $\ell$,
  so that quantum mechanics with highly-entangled quantum states can emerge in the large length limit $\ell \to \infty$.

\begin{figure}
  \centering
  \includegraphics[width=\columnwidth]{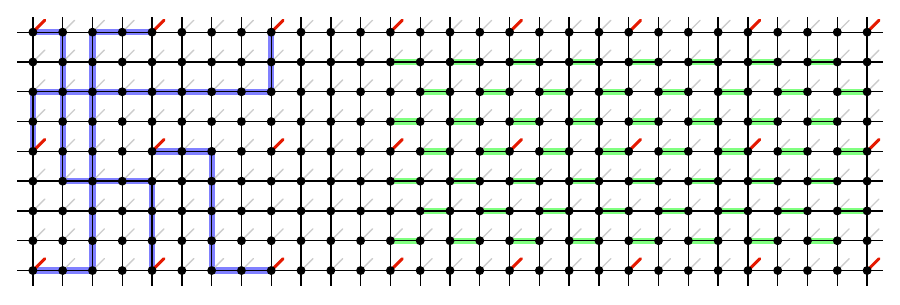} \\
  \caption{%
    A PEPS with physical qubits (dangling red lines) separated by many
      unimportant qubits (dangling gray lines), which we consider unimportant after coarse-graining.
    Black dots represent tensors which are contracted along the black lines.
    We see that the PEPS can easily represent many well-separated by Bell states
      by threading their entanglement along the blue lines.
    Furthermore, unitary circuits can be embedded into the PEPS
      (with unitary gates encoded in tensor pairs highlighted in green)
      to encode high-complexity states.
  }\label{fig:PEPS}
\end{figure}

On the other hand, PEPS do appear to be possibly powerful enough to represent wavefunctions
  that are sufficiently entangled to be compatible with current experimental observations.
For example, the mutual information bound for PEPS diverges linearly with the length scale: $I < O(\ell \log \chi)$.
Furthermore, due to their unconstrained nature,
  unitary circuits can easily be embedded into a PEPS in order to represent high-complexity wavefunctions.
This is especially true if the PEPS exists in extra spatial dimensions and/or
  if physical qubits are separated by large distances in the lattice of tensors,
  as in \figref{fig:PEPS}.

However, we are not aware of any time-evolving PEPS algorithms that could be written as a local classical model.
For example, PEPS update algorithms typically require calculating environment tensors,
  which involve a non-local contraction of the entire tensor network,
  for which non-local approximate algorithms are necessary \cite{AlgorithmsPEPS,GradientPEPS}.

However, the biggest challenge in formulating an EmQM model using a tensor network algorithm is that so far,
  no (local or nonlocal) time-evolving tensor network algorithm has been shown to be robust enough to create high-complexity states
  (\emph{e.g.} those created during the quantum advantage experiments \cite{GoogleSupremacy,ZuchongzhiAdvantage,ZuchongzhiAdvantage2,PhotonAdvantage})
  with fixed bond dimension.
With sufficiently large bond dimension, most tensor network algorithms can simulate essentially anything.
However in EmQM, we would like to imagine that the bond dimension is fixed and not too large.
Instead, it is the distance between physical qubits and time over which the dynamics occurs that is large,
  and it is over these extremely large scales that perhaps quantum mechanics could emerge.

\newpage
\section{Bell's Theorem and Tensor/Neural Networks}
\label{app:Bell}

Bell's theorem \cite{BellTheorem} and its conceptually simpler\footnote{%
    The GHZ variant is simpler in the sense that every measurement outcome is fixed according to quantum mechanics
      and simultaneously inconsistent with a local hidden variable theory.
    In contrast, Bell's original theorem only establishes a probabilistic Bell inequality.
    However, the GHZ variant involves three qubits instead of two.}
  GHZ variant \cite{BellGHZ,MerminBellGHZ,PhotonBellGHZ} show that
  local hidden variable theories are incompatible with quantum mechanics
  under the assumption that information can not travel faster than a certain speed.
In \appref{app:GHZ}, we review the GHZ variant.
In \appref{app:realism}, we then discuss why tensor networks and neural networks
  are not local hidden variable theories [spoiler: \eqsref{eq:realism} and \eqref{eq:local realism} are not satisfied]
  and can therefore bypass Bell's theorem (and its variants).
Other possible ways for EmQM to avoid Bell's theorem include the superdeterminism \cite{superdeterminism} loophole \cite{bellLoopholes}
  and the possibility that information in the underlying theory could travel faster than light,
  which could occur if the speed of light is also an emergent phenomena \cite{WenEmergent}.

\subsection{Review: GHZ Variant of Bell's Theorem}
\label{app:GHZ}

Writing the 3-qubit GHZ state in the ZZZ, YYX, and XXX basis yields:
\begin{align}
  \ket{\psi} &= \tfrac{1}{\sqrt{2}} \ket{+_z +_z +_z} + \tfrac{1}{\sqrt{2}} \ket{-_z -_z -_z} & ZZZ \text{ basis} \nonumber\\
             &= \tfrac{1}{2} \ket{-_y -_y -_x} + \tfrac{1}{2} \ket{-_y +_y +_x}
              + \tfrac{1}{2} \ket{+_y -_y +_x} + \tfrac{1}{2} \ket{+_y +_y -_x} & YYX \text{ basis} \label{eq:GHZ}\\
             &= \tfrac{1}{2} \ket{+_x +_x +_x} + \tfrac{1}{2} \ket{+_x -_x -_x}
              + \tfrac{1}{2} \ket{-_x +_x -_x} + \tfrac{1}{2} \ket{-_x -_x +_x} & XXX \text{ basis} \nonumber
\end{align}
$\ket{\psi}$ takes a similar permuted form in the YXY and XYY basis.
We use the following notation for the qubit states
\begin{gather}
\begin{aligned}
  \ket{\pm_x} &= \tfrac{1}{\sqrt{2}} \ket{+_z} \pm \tfrac{1}{\sqrt{2}} \ket{-_z} \\
  \ket{\pm_y} &= \tfrac{1}{\sqrt{2}} \ket{+_z} \pm \tfrac{i}{\sqrt{2}} \ket{-_z} \\
\end{aligned}
\end{gather}
  which are eigenstates of the Pauli operators:
  $\hat{\sigma}^\mu \ket{\pm_\mu} = \pm \ket{\pm_\mu}$.
In this Appendix, operators will always be denoted with a hat.
Note that the GHZ state has a fixed measurement outcome for the $\hat{Y}_1 \hat{Y}_2 \hat{X}_3$ and $\hat{X}_1 \hat{X}_2 \hat{X}_3$ operators
  (along with permutations)
\begin{equation}
\begin{aligned}
\braket{\psi|\hat{Y}_1 \hat{Y}_2 \hat{X}_3|\psi} &= -1\\
\braket{\psi|\hat{X}_1 \hat{X}_2 \hat{X}_3|\psi} &= +1
\end{aligned} \label{eq:quantum exp}
\end{equation}
$\hat{X}_i$ denotes the $\hat{\sigma}^x$ Pauli operator acting on the $i^\text{th}$ qubit,
  and similar for $\hat{Y}_i$ and $\hat{Z}_i$.

In a hidden variable theory\footnote{%
    Other definitions of hidden variable theories exist in the literature \cite{AaronsonHiddenVariable}.
    We are reviewing the definition as used in Bell's theorem.},
  measurement outcomes only depend on a hidden classical variable $h$,
  which has a probability density $P(h)$.
The measurement outcome due to measuring a qubit $i$ in the X, Y, or Z basis can then be written as:
\begin{align}
  X_i(h) &= \pm1 \nonumber\\
  Y_i(h) &= \pm1 \label{eq:realism}\\
  Z_i(h) &= \pm1 \nonumber
\end{align}
Therefore, the quantum expectation values reduce to classical expectation values:
\begin{align}
  \braket{\psi|\hat{X}_i|\psi} &= \int_h P(h) \, X_i(h) \nonumber\\
  \braket{\psi|\hat{Y}_i|\psi} &= \int_h P(h) \, Y_i(h) \\
  \braket{\psi|\hat{Z}_i|\psi} &= \int_h P(h) \, Z_i(h) \nonumber
\end{align}
In a \emph{local} hidden variable theory, measuring qubit $i$ can not affect the measurement outcome of qubit $j$
  if qubit $j$ is far away and measured immediately after qubit $i$.
Therefore, for space-like separated measurements in a local hidden variable theory, the multi-point quantum expectation values
  also reduce to analogous classical expectation values; in particular:
\begin{equation}
\begin{aligned}
  \braket{\psi|\hat{Y}_1 \hat{Y}_2 \hat{X}_3|\psi} &= \int_h P(h) \, Y_1(h) Y_2(h) X_3(h) \\
  \braket{\psi|\hat{X}_1 \hat{X}_2 \hat{X}_3|\psi} &= \int_h P(h) \, X_1(h) X_2(h) X_3(h)
\end{aligned} \label{eq:local realism}
\end{equation}

To obtain a contradiction with quantum theory, note that
  \eqsref{eq:quantum exp},
  \eqref{eq:realism},
  \eqref{eq:local realism},
  and permutations imply that
\begin{align}
  Y_1(h) Y_2(h) X_3(h) &= Y_1(h) X_2(h) Y_3(h) = X_1(h) Y_2(h) Y_3(h) = -1 \label{eq:local YYZ} \\
  X_1(h) X_2(h) X_3(h) &= +1 \label{eq:local XXX}
\end{align}
for all hidden variables $h$ (with $P(h) \neq 0$).
However, $X_1(h) X_2(h) X_3(h)$ can be rewritten using \eqnref{eq:local YYZ}
\begin{gather}
\begin{aligned}
  X_1(h) X_2(h) X_3(h) &= [Y_1(h) Y_2(h) X_3(h)] \, [Y_1(h) X_2(h) Y_3(h)] \, [X_1(h) Y_2(h) Y_3(h)] \\
    &= [-1] [-1] [-1] = -1
\end{aligned} \label{eq:local XXX'}
\end{gather}
since $Y_i(h)^2 = 1$.
The above \eqnref{eq:local XXX'} contradicts \eqnref{eq:local XXX}.
This shows that quantum mechanics can not be described by a local hidden variable theory.

\subsection{Tensor and Neural Networks}
\label{app:realism}

Tensor networks and neural networks might seem like hidden variable theories.
However, neither are local hidden variable theories
  and are therefore not ruled out by Bell experiments.
The primary issue is that observables are not well-defined in terms of hidden variables, as in \eqnref{eq:realism},
  in tensor or neural networks.

Furthermore, in the limit of infinite bond dimension or number of training parameters,
  tensor and neural networks can simulate quantum mechanics exactly \cite{CiracTensorRev,StoudenmireLimits,PollmannNeural},
  which implies that Bell's theorem must not be violated in that limit.
If an exact tensor or neural network (or other EmQM model in an exact limit) were to simulate a Bell experiment,
  then there would be no sense of local realism.
That this, after the observers measured their qubits,
  the EmQM model would be encoding a wavefunction that consists of a superposition of measurement outcomes,
  as predicted by Schr\"{o}dinger's equation.
As emphasized in the many-worlds interpretation of quantum mechanics,
  the wavefunction would then approximately consist of multiple ``branches'' that evolve independently in superposition
  since they are extremely unlikely to interfere with each other\footnote{%
    This assumes that the observers (including the surrounding environment) consist of many chaotic degrees of freedom, which is certainly true for human observers.}.
In an actual non-exact EmQM model,
  only a limited number of wavefunction branches could be retained.
Therefore, \emph{after} the measurements took place,
  an EmQM model would eventually be forced to randomly drop wavefunction branches
  (especially if many Bell measurements are performed).
As long as only valid branches are retained and branches are dropped with the correct probability,
  the Bell experiments would not notice deviations from quantum mechanics.
In a matrix product state (MPS) tensor network,
  dropping a branch could be achieved by a special random (or pseudo-random) truncation of a bond dimension.

\subsubsection{GHZ Example}

For example, the 3-qubit GHZ state from \eqnref{eq:GHZ} can easily be encoded in a
  matrix product state (MPS) tensor network in the ZZZ basis:
\begin{align}
  \ket{\psi} &= \sum_{a_1,a_2,a_3,\alpha,\beta=0,1} A_{a_1,\alpha} \, B_{\alpha,a_2,\beta} \, C_{\beta,a_3}
                \ket{a_1a_2a_3} \\
  A_{a_1,\alpha} &= \delta_{a_1,\alpha} \nonumber\\
  B_{\alpha,a_2,\beta} &= 2^{-1/2} \delta_{\alpha,a_2,\beta} \\
  C_{\beta,a_3} &= \delta_{\beta,a_3} \nonumber
\end{align}
  where $(a_1,a_2,a_3)$ index states in the Z basis ($\ket{0} = \ket{+_z}$ and $\ket{1} = \ket{-_z}$)
  and $(\alpha,\beta)$ are virtual indices.
$\delta$ is the Kronecker delta, with $\delta_{a,b,c} = 1$ if $a=b=c$ else $\delta_{a,b,c} = 0$.

One could attempt to map the above tensor network to a hidden variable theory by defining the hidden variable as
  the physical and virtual indices: $h=(a_1,a_2,a_3,\alpha,\beta)$.
Then the Z-basis measurements obey the analogs of Eqs.\,(\ref{eq:realism}--\ref{eq:local realism}) with
  $P(h) = \frac{1}{2} \delta_{a_1,a_2,a_3,\alpha,\beta}$ and $Z_i(h) = (-1)^{a_i}$.
However, there is no choice of $X_i(h)$ and $Y_i(h)$ that is also consistent with Eqs.\,(\ref{eq:realism}--\ref{eq:local realism}).\footnote{%
  One might alternatively imagine adding $\alpha'$ and $\beta'$ to the hidden variable in order to account for
    virtual indices for a bra state.
  However, this does not help turn the tensor network into a local hidden variable theory.}

Neural networks also fail to be a local hidden variable theory in a similar way.

\end{document}